\documentclass[fontsize=11pt,a4paper]{article}
\usepackage[T1]{fontenc}
\usepackage[utf8]{inputenc}
\usepackage[english]{babel}
\usepackage{graphicx}
\usepackage[a4paper]{geometry}
\usepackage{lmodern}
\usepackage{lineno}
\usepackage{color}

\ifdefined\REVIEW
   \linenumbers
   
\fi

\usepackage{hyperref}
\hypersetup{
    colorlinks=true,
    linkcolor=blue,
    citecolor=blue,
    urlcolor=blue}
\usepackage{amsmath} 
\usepackage{amsfonts} 
\usepackage{amssymb}
\usepackage{cite}
\usepackage{physics}

\newcommand{\eq}[1]{(\ref{eq:#1})}
\newcommand{\Eq}[1]{Eq.\,\eq{#1}}

\newcommand{\Fig}[1]{Fig.~\ref{fig:#1}}
\newcommand{\fig}[1]{\ref{fig:#1}}

\title{Universal Dynamics at the Lowest Temperatures}

\author{
        Ido Siovitz, Philipp Heinen, Niklas Rasch, Stefan Lannig, Yannick Deller, \\
        Helmut Strobel, 
        Markus Oberthaler, and Thomas Gasenzer \\[1.5ex]
        \textit{Kirchhoff-Institut f\"ur Physik,
        Im Neuenheimer Feld 227,
        Universit\"at Heidelberg, Germany}
}

\date{}
\begin{document}
\maketitle

%
%

\begin{abstract}
High-performance graphical processing units (GPU) are used for the repeated parallelised propagation of non-linear partial differential equations on large spatio-temporal grids.
The main challenge results as a combination of the requirement of large grids for exploring scaling over several orders of magnitude, both in space and time, and the need for high statistics in averaging over many runs, in computing correlation functions for highly fluctuating quantum many-body states.
With our simulations, we explore the dynamics of complex quantum systems far from equilibrium, with the aim of classifying their universal characteristics such as scaling exponents near non-thermal fixed points.
Our results are strongly relevant for the development of synthetic quantum systems when exploring the respective physics in the laboratory. 
\end{abstract}

\section{Introduction}\label{introduction}

The characterisation of complex physical structures, in particular when being far from equilibrium and exhibiting strong correlations, poses a long-standing challenge to physics. 
Of particular interest are universal aspects of equilibrium as well as non-equilibrium systems, which imply that experiments on one system can be relevant for the understanding of entirely different ones.
An increasing interest lies also on the question to what extent complex structures could be used for developing new technologies for physical computation.
A primary task is to develop theoretical and experimental tools for precisely controlling many-particle systems and observing their dynamics. 
Basic strategies comprise extending the range of predictive methods by studying prototypical models numerically. 
These models allow mutual benchmarks in terms of mathematical analysis, numerical simulations, as well as measurements on well-controlled synthetic quantum systems, in our context ultracold atomic gases.
Such systems typically comprise a few thousand to million atoms cooled to temperatures in the nano Kelvin regime, where they exhibit quantum properties and in particular undergo a transition to a so-called Bose-Einstein condensate, forming a nearly incompressible superfluid that flows without friction. 
 
During recent years, we have developed a strong focus on far-from-equilibrium dynamics as well as on strong correlations in such systems. 
Our specific research includes the development and application of parallelised computing techniques, which ensure reaching the necessary statistical precision and system sizes for the characterisation of the considered dynamical phenomena and properties. 
Exemplary topics include the classification of universal quantum dynamics near non-thermal fixed points of the time evolution, the precise characterisation of strongly correlated equilibrium and non-equilibrium states, as well as long-time evolution and the approach to thermal equilibrium.

\section{Simulations of ultracold superfluid quantum gases}\label{theory}

In the following we give a brief introduction to the main aspects of universal dynamics near a non-thermal fixed point, for the example of a three-component ultracold atomic gas and discuss the challenges posed by statistical simulations of the highly fluctuating and correlated system, which we perform in a  highly parallelised manner on clusters of graphical processing units (GPU). 

\subsection{Universal dynamics far from equilibrium}

Quantum many-body systems driven far out of equilibrium can show universal dynamical behaviour \cite{%
Prufer:2018hto,
Eigen2018a.arXiv180509802E,
Erne:2018gmz,
Gauthier2019a.Science.364.1264,
Johnstone2019a.Science.364.1267,
Navon2018a.doi:10.1126/science.aau6103,
Glidden:2020qmu,
GarciaOrozco2021a.PhysRevA.106.023314,
Berges:2008wm,
Schole:2012kt,
Orioli:2015dxa,
Williamson2016a.PhysRevLett.116.025301,
Karl2017b.NJP19.093014,
Walz:2017ffj.PhysRevD.97.116011,
Chantesana:2018qsb.PhysRevA.99.043620,
Mikheev:2018adp,
Schmied:2018upn.PhysRevLett.122.170404,
Mazeliauskas:2018yef,
Schmied:2018mte,
Schmied:2018osf.PhysRevA.99.033611,
Gao2020a.PhysRevLett.124.040403,
Wheeler2021a.EPL135.30004,
Gresista:2021qqa,
RodriguezNieva2021a.arXiv210600023R,
Heinen:2022rew%
}. 
This includes, e.g., systems suddenly quenched to a non-equilibrium state and relaxing back to equilibrium. 
Universality typically means that correlations between fields evaluated at in general different space (and time) points show a simple form independent of the particular details of the system, defined only by symmetries of the underlying model \cite{Schmied:2018mte}. 
They are commonly characterised by power laws in momentum space and frequency, and also their change in time reduces to a rescaling defined by a few universal exponents only. 
Examples include the evolution of ensembles of topological excitations and their quantum turbulent dynamics, phase-ordering processes, and other phenomena described by so far unknown non-thermal fixed points \cite{%
Berges:2008wm,
Orioli:2015dxa,
Schmied:2018mte%
}.
The goal is to understand which types of such dynamics there are, i.e., to find and characterise the universality classes of non-thermal fixed points, in order to learn about possible dynamics of similar type in entirely different systems, ranging, e.g., from structure formation in the early universe, following the big bang, via astrophysical objects such as neutron stars, to dynamics of ocean waves and microscopic phenomena in the solid state.

%
\begin{figure}[t]
\begin{center}
\includegraphics[width=0.9\textwidth]{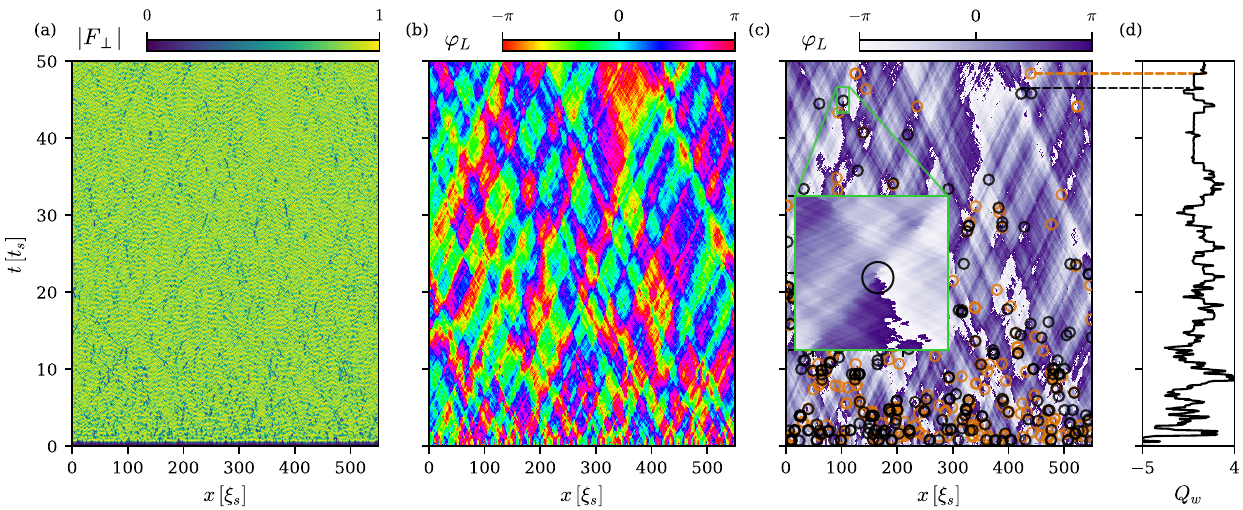}
\end{center}
\caption{\label{fig:RealSpaceEvoSpin} 
Space-time evolution of a spin $\mathbf{F}_{\perp}(t,x)=(F_{x},F_{y})$ of length $|F_{\perp}| = ({F_{x}^{2}+F_{y}^{2}})^{1/2}$ and orientation angle $\varphi_{\mathrm{L}}=\mathrm{arg}(F_x+\mathrm{i}F_{y})$ in the $F_{x}$-$F_{y}$-plane. 
Starting with an equilibrated but fluctuating system and suddenly changing some parameter, here an external magnetic field, introduces short-wave-length excitations which subsequently lead to the formation of textures in the transversal spin after a few characteristic time scales $t_\mathrm{s}$. 
These spin textures, i.e., domains of different orientation (panel b), with overall constant spin length (a) are separated by kink-like defects and travel with a roughly uniform velocity in either direction.  
The size of the spin textures grows in time and goes together with a decreasing frequency of the occurrence of vortex-type defects in space and time (c) each of which changes the overall winding number  $Q_{w}(t)=\int\mathrm{d}x \,\partial_{x}\varphi_\mathrm{L}(t,x)$ of the Larmor phase $\varphi_\mathrm{L}$ (d) along the system with a periodic boundary condition.
See \cite{Schmied:2018osf.PhysRevA.99.033611,Siovitz2023a} for more details.
}
\end{figure}
%
Consider, for instance, a one-dimensional Bose-Einstein condensate, which consists of bosonic atoms that can move along the $x$ direction, and which can be in three different internal electronic states.
Its state can be described by a spin vector field $\mathbf{F}(t,x)=(F_{x},F_{y},F_{z})(t,x)$, which encodes the local density of atoms and the mean orientation of $\mathbf{F}$ representing the atoms' three-component quantum mechanical spin-1 state.

\Fig{RealSpaceEvoSpin} shows an example of such a field $F_{\perp}\equiv F_{x}+\mathrm{i}F_{y}= |F_{\perp}| \exp \left \{{\mathrm{i} \varphi_\mathrm{L}}\right \}$ separated into spin length $|F_{\perp}|$ and orientation angle $\varphi_\mathrm{L}$ between the vector $(F_{x},F_{y})$ and the $F_{x}$-axis \cite{Schmied:2018osf.PhysRevA.99.033611,Siovitz2023a},
initially fluctuating weakly about a vanishing mean $\langle\mathbf{F}(0,x)\rangle\equiv0$.
Both the length and orientation grow and thereby fluctuate in time.
While (a) shows that the length rather quickly assumes a fairly constant value, (b) exhibits large patches of equal orientation, separated by sharp jumps in the angle. 
The jumps propagate with a fairly constant velocity forward and backward along the spatial coordinate, which is here measured in units of the so-called spin coherence length $\xi_\mathrm{s}$.
On average, the interval lengths of co-alignment of the spin slowly increase.
Such a phenomenon is typically called domain coarsening and known to be associated with power-law evolution \cite{Bray1994a.AdvPhys.43.357,Cugliandolo2014arXiv1412.0855C}.
Hence, there is a correlation length scale $\ell_{\Lambda}$ governing the size of the patches of equal colour, which grows in time as $\ell_\Lambda(t)\sim t^{\beta}$, with the exponent being $\beta\simeq1/4$ as found in \cite{Schmied:2018osf.PhysRevA.99.033611}.
More recent investigations indicate that the coarsening process goes along with the appearance of vortex-like defects in space and time (\Fig{RealSpaceEvoSpin}c) \cite{Siovitz2023a}, which give rise to jumps of the winding number $Q_{w}=\int\mathrm{d}x \,\partial_{x}\varphi_\mathrm{L}(x)$ and which can be associated with caustics appearing in the chaotic wave propagation, similar to rogue waves on ocean surfaces \cite{Onorato2013a.PhysRep528.47} (\Fig{RealSpaceEvoSpin}d).

\begin{figure}[t]
\begin{center}
\includegraphics[width=0.9\textwidth]{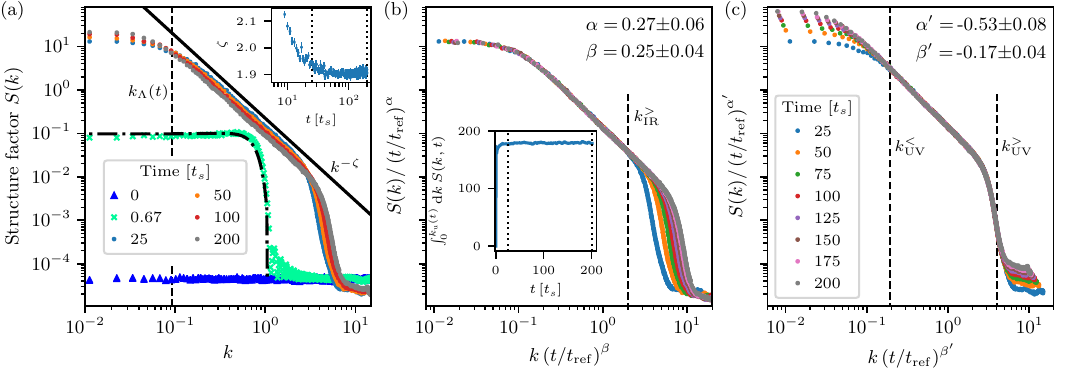}
\end{center}
\caption{\label{fig:StructureFactor}
Scaling evolution of the structure factor $S(k,t)$. 
(a) Following a fast initial growth of modes up to a certain maximum, the form of $S(k,t)$ approaches a universal shape, which only slowly evolves and rescales in time, \Eq{StructureFactor}, separately in the 
(b) IR regime of momenta, towards lower $p$, with $\beta\simeq0.25$, $\alpha\simeq0.27$, thereby conserving the integral under the curve (see inset), and in the 
(c) UV, where a different set of (negative) exponents implies that the scaling is towards higher momenta.
The shape of the curve becomes universal, in particular also exhibiting a constant power-law fall off $\sim k^{-\zeta}$ (see inset of panel a).
The rescaling of the IR scale $k_{\Lambda}(t)\sim t^{-\beta}$ reflects the coarsening seen in \Fig{RealSpaceEvoSpin}, i.e., a self-similar growth of the spin-orientation pattern characterised by a correlation length $\ell_{\Lambda}\sim k_{\Lambda}^{-1}$.
The exponents together with the scaling form $f_\mathrm{s}$ define the universality class the underlying non-thermal fixed point belongs to.
Figure taken from Ref.~\cite{Schmied:2018osf.PhysRevA.99.033611}.
}
\end{figure}

Taking averages over many such evolutions $\mathbf{F}(t,x)$, starting from said randomly fluctuating initial configurations, we typically compute a so-called structure factor, e.g., $S(t,k)=\langle F_{\perp}(t,k)^{*}F_{\perp}(t,k)\rangle$, of the Fourier transform $F_{\perp}(t,k)$ to momenta $k$, of the spin orientation $F_{\perp}(t,x)=(F_{x}+\mathrm{i} F_{y})(t,x)$ in the $F_{x}$-$F_{y}$-plane.
This quantity encodes the time evolving spectrum of excitations exemplarily shown in \Fig{RealSpaceEvoSpin}.
\Fig{StructureFactor} depicts the resulting averaged structure factor: 
Starting from a constant distribution over all momenta (blue triangles), during the early evolution, a strong buildup of excitations, up to a maximum momentum $k_{Q}\simeq 1$ set by the quench, is seen (green crosses), which, at late times, goes over into a universal form with a plateau in the infrared (IR), a power-law fall off $\sim k^{-\zeta}$ at larger momenta, and a steep fall-off at the ultraviolet (UV) end.
This distribution shows scaling evolution according to 
\begin{equation}
    S(t,k) = (t/t_\mathrm{ref})^\alpha f_\mathrm{s}([t/t_\mathrm{ref}]^{\,\beta} k)
    \,.
    \label{eq:StructureFactor}
\end{equation}
Here $f_\mathrm{s}$ represents the universal scaling function, i.e., the shape of $S(t,k)$ (considered separately in the IR, $k<k^{>}_\mathrm{IR}$, and UV, $k^{<}_\mathrm{UV}<k<k^{>}_\mathrm{UV}$), which depends only on the momentum $k$, $t_\mathrm{ref}$ is a reference time within the range of times where scaling prevails, and the scaling exponents characterise the IR (panel b) or UV (c) rescaling, with values indicated inside the respective panels.
In the IR, they are related by $\alpha=d\beta$, ensuring the momentum integral over $S(t,k)$ to be conserved in time.

\subsection{Highly parallelised statistical simulations of the quantum field dynamics}

The systems' dynamics is subject to non-linear coupled differential equations for multi-component quantum fields $\psi_{m}(t,\mathbf{x})$, in one or more spatial dimensions, which for the above 3-component system take the form
\begin{align}
    \mathrm{i}\hbar\partial_t\mqty(\psi_{1}\\ \psi_{0}\\ \psi_{-1})
     = \left[-\frac{\hbar^2}{2m}\Delta +q+(c_{0}+c_{1})\rho
     +c_1 
	\mqty(
	 -2|\psi_{-1}|^{2} & \psi_{-1}^*\psi_0 & 0 \\
	\psi_{-1}\psi_0^* & -|\psi_{0}|^{2}-q/c_{1} & \psi_0^*\psi_1 \\
	0 & \psi_0\psi_1^* & -2|\psi_{1}|^{2}) 
     \right]\mqty(\psi_{1}\\ \psi_{0}\\ \psi_{-1})
     \,,
     \label{eq:GPE}
\end{align}
where $\hbar$, $m$, $q$, and $c_{0,1}$ are real-valued constants and the fields $\psi_{m}(t,x)$ are complex and define the local density of atoms $\rho=\sum_{m=0,\pm1}|\psi_{m}|^2$.
As in quantum mechanics, the three-component field defines a spinor, which encodes the mean orientation and strength of the angular momentum vector $\mathbf{F}$, for which examples were shown in Figs.~\fig{RealSpaceEvoSpin} and \fig{StructureFactor}.
To solve the equations, we use a split-step Fourier method, in which the operator in \eq{GPE} in square brackets is split into three parts, each of which we use to propagate in one time step from $t_{n}$ to $t_{n+1}$ separately: 
The kinetic part involving the Laplacian $\Delta=\sum_{i=1}^{d}\partial_{i}^{2}$, which is readily diagonal in momentum space;
the diagonal interaction part, which includes all terms except those in the off-diagonal elements of the matrix,
and the off-diagonal interaction part, which accounts for component mixing.
The first step involves two Fourier transforms $\mathcal{F}$, with a phase evolution over time step $\Delta t=t_{n+1}-t_{n}$ sandwiched in between, while the second and third steps involve a phase and matrix multiplication directly in position space,
\begin{align}
	\vec{\psi}'(x_j,t_{n+1}) 
	&\equiv \mqty(\psi_{1}, \psi_{0}, \psi_{-1})^{T}(x_j,t_{n+1})
	=\mathcal{F}^{-1}\left\{e^{-\mathrm{i}\Delta t |k|^2/2} \mathcal{F}\left[ \vec{\psi}(x_j,t_n)\right](k)\right\}
	\,,\\    
	\vec{\psi}''(x_{j},t_{n+1}) 
	&= \exp\left\{-\mathrm{i}\left[V_m(x_j) + f_m(|\psi_0(x_j,t_n)|^2,|\psi_1(x_j,t_n)|^2,|\psi_{-1}(x_j,t_n)|^2)\right]\Delta t\right\}
	\vec{\psi}'(x_{j},t_{n+1})
	\,,
	\\    
	\vec{\psi}(x_j,t_{n+1}) 
	&=
	\frac{1}{\lambda^{2}}\mqty(
	{|a|^2\cos(\lambda c_1 \Delta t) + |b|^2} & 
	-\mathrm{i}a{\lambda}\,{\sin(\lambda c_1 \Delta t)} & 
	ab {\cos(\lambda c_1 \Delta t)-1} \\
	-\mathrm{i}a^*{\lambda}\,{\sin(\lambda c_1 \Delta t)} & 
	{\lambda}^{2}\cos(\lambda c_1 \Delta t) & 
	-\mathrm{i}b{\lambda}\,{\sin(\lambda c_1 \Delta t)} \\
	a^*b^* {\cos(\lambda c_1 \Delta t)-1} & 
	-\mathrm{i}b^*{\lambda}\,{\sin(\lambda c_1 \Delta t)} & 
	{|a|^2 + |b|	^2\cos(\lambda c_1 \Delta t)}) 
	\vec{\psi}''(x_j,t_{n+1})
	\,.
\end{align}
Here, $V_{m}$ and $f_{m}$ include the potential and diagonal non-linear parts of the operator in \eq{GPE} in square brackets, and $\lambda= \sqrt{|a|^2 + |b|^2}$, with $a = \left[\psi_{-1}^*\psi_0 + \tilde{\psi}_{-1}\tilde{\psi}_0\right]/2$, 
$b = \left[\psi_{0}^*\psi_1 + \tilde{\psi}_{0}\tilde{\psi}_1\right]/2$, and ${\tilde{\psi}_{m}} = {\psi}_{m} - \mathrm{i}S_{mn}\psi_{n}$ with $S$ being the off-diagonal part of the matrix in \eq{GPE}.

The above propagation needs to be performed on sufficiently large spatial and temporal grids, many times in sequence, starting from slightly fluctuating initial conditions, in order to accumulate sufficient statistics for evaluating the average correlations such as shown in \Fig{StructureFactor} and to take into account quantum fluctuations.
This requires computational resources exceeding by far local PC clusters.
As an example, in our 1D simulations, we chose a spatial grid of 4096 lattice points, each of which needs to hold  3 complex numbers, which results in a $3\times2\times4096= 24$\,k grid of at least double-precision numbers.
We start as many runs simultaneously on each GPU (Nvidia V100 und A100), as its RAM captures, and one such total run for the example case described above is over typically ca.~$1.5$\,M time steps. 
It lasts between $120$ and $150\,$s wall time, and thus it takes $\sim11\,$hrs for $250$ runs, or for $1000$ runs on a machine with 4 GPUs, typically needed for achieving well averaged distributions. 
As compared with computations on CPU clusters alone, the described combination has an advantage of roughly a factor of 10 for our purposes.
2D and 3D systems require smaller grids and less runs to average well, so far up to $256^{3}\times3\times2$ in three dimensions, with larger grids being possible for systems with less internal components. 
All data evaluation is done in parallel on CPUs, and the code has been developed in C++ with CUDA and OpenMP parallelisation.

\section{Discussion and Conclusions}
\label{sec:conclusions}

Architectures of high-performance graphical processing units (GPU) are used for the repeated parallelised propagation of non-linear partial differential equations on large spatio-temporal grids.
The main challenge consists of a combination of several bottleneck-like limitations:
On the one side, studies of universal dynamics require the evaluation of the time-evolving field configurations within large volumes \emph{and} with high resolution, in order to span as many orders of magnitude in spatial extent as possible.
This requirement results from the goal to detect self-similar scaling behaviour with sufficient precision.
On the other hand, the considered dynamics far from thermal equilibrium gives rise to strong fluctuations and catastrophe-type events such as rogue waves and caustics, as well as non-linear excitations like topological defects.
As a result, a sufficiently high temporal resolution is needed to capture their short-time characteristics, while typically only long evolution times give rise to scaling behaviour.
Moreover, due to the strong fluctuations, many runs, starting from slightly different initial conditions, are required in order to achieve sufficient averaging statistics in evaluating correlation functions, i.e., moments of the underlying probability distributions. 

The exemplary results shown here demonstrate the power of the parallelised evaluation of our simulations, which allows a speedup by up to a factor 10 as compared to standard OpenMP parallelisation on CPUs. 
So far, this has been considered to be of higher relevance than the limitations set by the RAM size of available high-end GPUs.
Possible future extensions include a parallelisation of the lattice representation of a single system over several GPUs.
Such an extension is technically feasible but represents a severe limitation for the type of split-step Fourier propagation we employ in propagating the differential equations. 
Techniques developed for the grid-based numerical solution of partial differential equations such as \textsc{DUNE} \cite{Bastian2021aCompMathApp81.75} could help overcoming the RAM limitations but likely require to avoid split-step Fourier in order to reduce data exchange between different GPU units.
This poses a challenge for the evolution of PDEs of the non-linear Schr\"odinger type, which involve topological defect solutions which are most easily captured within the chosen approach.

Even harder challenges prevail in computing full quantum fluctuation properties of many-body systems, both in and, even more so, out of equilibrium.
Technically similar techniques can be used, when extended to twice as large configuration spaces, to determine properties of highly correlated quantum systems, e.g., near equilibrium phase transitions \cite{Heinen2022a.PhysRevA106.063308.complex}.
In summary, the high-performance computing facilities provided by {bw$|$HPC} represent a decisive element in efforts to explore, model, and understand fundamental characteristics of the physics of complex quantum systems, in view of near-future developments in controlling such systems in a tailored way in the laboratory.

\section{Acknowledgements}

The authors acknowledge support 
by the German Research Foundation (Deutsche Forschungsgemeinschaft, DFG), through 
SFB 1225 ISOQUANT (Project-ID 273811115), 
grant GA677/10-1, 
and under Germany's Excellence Strategy -- EXC 2181/1 -- 390900948 (the Heidelberg STRUCTURES Excellence Cluster), 
by the state of Baden-W{\"u}rttemberg through bwHPC and DFG through
grants INST 35/1134-1 FUGG, INST 35/1503-1 FUGG, INST 35/1597-1 FUGG, and 40/575-1 FUGG.



\end{document}